%% file: main.tex
\documentclass[final]{cvpr}
\pdfoutput=1
\usepackage{times}
\usepackage{epsfig}
\usepackage{placeins}
\usepackage{multicol, caption}
\usepackage{docmute}
\usepackage{graphicx}
\usepackage{amsmath}
\usepackage{afterpage}
\usepackage{amssymb}
\usepackage{booktabs}

\usepackage{chapterbib}

\usepackage[ruled,noend]{algorithm2e}
\usepackage[dvipsnames]{xcolor}
\usepackage{import}

\def\HiLi{\leavevmode\rlap{\hbox to \hsize{\color{lightgray!50}\leaders\hrule height .8\baselineskip depth .5ex\hfill}}}

\SetCommentSty{mycommentfont}

\usepackage[pagebackref=true,breaklinks=true,colorlinks,bookmarks=false]{hyperref}

\begin{document}

\include{latex/mantissacam}

\setcounter{figure}{0}
\setcounter{section}{0}
\setcounter{equation}{0}

\include{latex/supplement}

\end{document}

%% file: latex/mantissacam.tex
\title{MantissaCam: Learning Snapshot High-dynamic-range Imaging with Perceptually-based In-pixel Irradiance Encoding}

\author{Haley M. So$^1$ $\quad$ Julien N.P. Martel$^1$ $\quad$ Piotr Dudek$^2$ $\quad$ Gordon Wetzstein$^1$\\
$^1$Stanford University $\quad$ $^2$The University of Manchester\\
{\tt\small \href{https://www.computationalimaging.org/publications/mantissacam/}{computationalimaging.org/publications/mantissacam/} }
}
\date{}
\maketitle

\begin{abstract}
The ability to image high-dynamic-range (HDR) scenes is crucial in many computer vision applications. The dynamic range of conventional sensors, however, is fundamentally limited by their well capacity, resulting in saturation of bright scene parts. To overcome this limitation, emerging sensors offer in-pixel processing capabilities to encode the incident irradiance. Among the most promising encoding schemes is modulo wrapping, which results in a computational photography problem where the HDR scene is computed by an irradiance unwrapping algorithm from the wrapped low-dynamic-range (LDR) sensor image. Here, we design a neural network--based algorithm that outperforms previous irradiance unwrapping methods and we design a perceptually inspired ``mantissa'' encoding scheme that more efficiently wraps an HDR scene into an LDR sensor. Combined with our reconstruction framework, MantissaCam achieves state-of-the-art results among modulo-type snapshot HDR imaging approaches. We demonstrate the efficacy of our method in simulation and show benefits of our algorithm on modulo images captured with a prototype implemented with a programmable sensor. 
\end{abstract}


\section{Introduction}
\label{sec:intro}
\input{latex/sections/introduction}

\section{Related Work}
\label{sec:related}
\input{latex/sections/related}

\section{Perceptually-based HDR Imaging}
\label{sec:method}
\input{latex/sections/method}

 \section{Experiments}
\label{sec:experiments}
\input{latex/sections/experiments}

\section{Discussion}
\label{sec:discussion}
\input{latex/sections/discussion}

\section*{Acknowledgements}
This project was in part supported by NSF Award 1839974, the NSF Graduate Research Fellowship, and a PECASE by the ARL.


%% file: latex/sections/introduction.tex
High Dynamic Range (HDR) imaging is crucial for a vast range of applications, including automotive vision systems~\cite{knoll2007hdr}, HDR display~\cite{Seetzen:2004}, and image processing~\cite{reinhard2010high,Banterle:2011}. When capturing natural scenes, which can have an extreme high dynamic range~\cite{reinhard2010high}, the level of detail is limited by the full well capacity and the quantization precision of the sensor. Unfortunately, the dynamic range offered by modern sensors is far smaller than that encountered in the wild~\cite{Ohta:2020}, making specialized sensors or computational photography approaches to HDR imaging necessary. 

\begin{figure}[!t]
\begin{center}
  \includegraphics[width=\columnwidth]{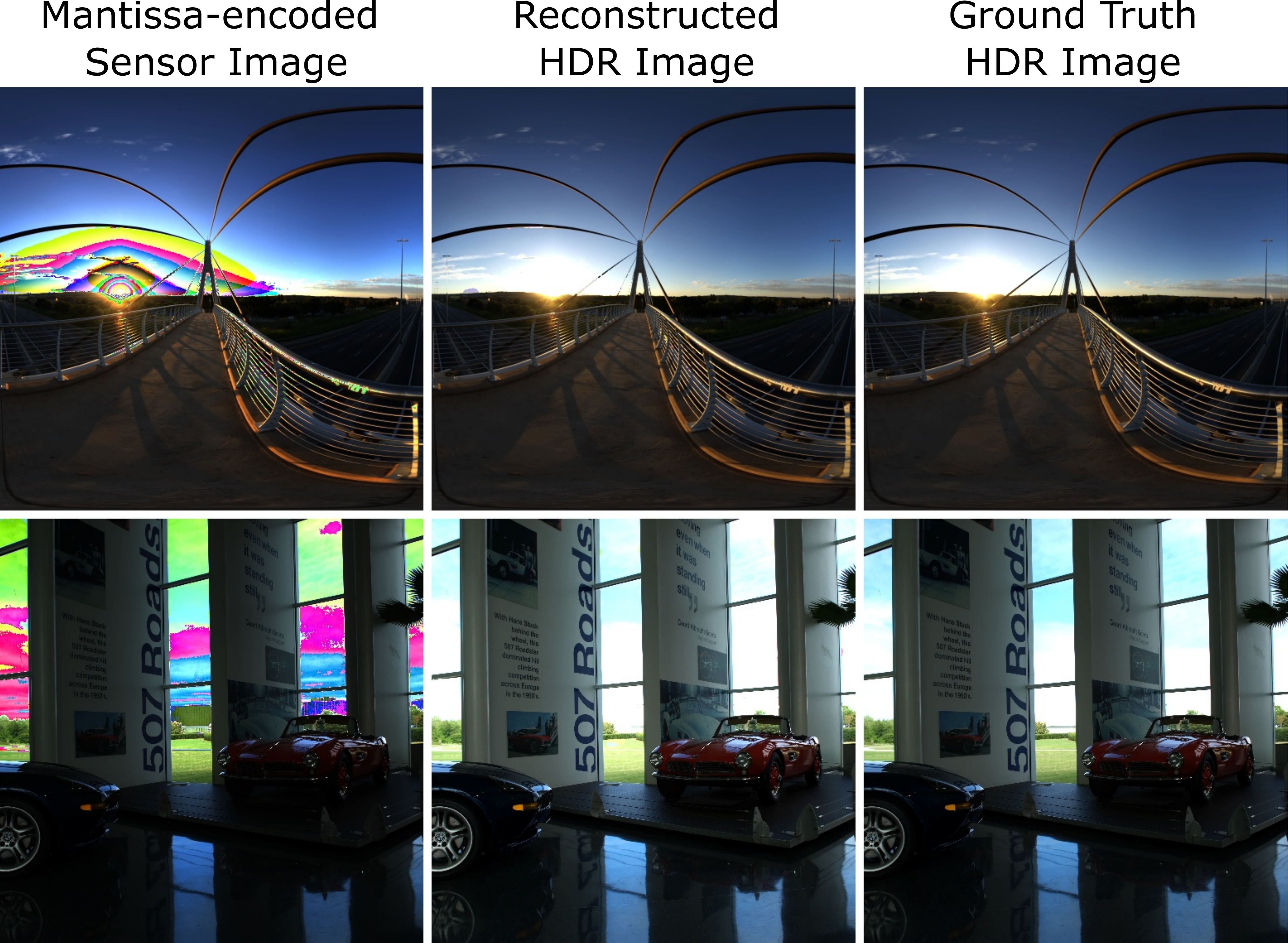}
\end{center}
  \caption{MantissaCam electronically encodes the irradiance incident on the sensor into an LDR image by wrapping the intensity in a perceptually inspired manner (left). The proposed reconstruction algorithm estimates the HDR scene from this LDR image (center) and achieves accurate reconstructions compared to the ground truth (right). }
	\label{fig:teaser}
\end{figure}

Among the many HDR imaging techniques proposed in the literature, exposure bracketing~\cite{Picard95onbeing,Debevec:1997,mertens2009exposure,Hasinoff:2010,gupta:2013,HDRplus} and temporally varying exposures~\cite{Kang:2003,Sen:2012:RPH,HDRVideo} can be successful, but fast motion introduces ghosting. Multi-sensor approaches~\cite{DBLP:journals/ijcv/AggarwalA04,McGuire:2007,Tocci:2011} can overcome this limitation, but are expensive, bulky, and difficult to calibrate. Existing snapshot HDR imaging approaches hallucinate saturated image detail using neural networks~\cite{marnerides2018expandnet,HDRCNN,Endo:2017,lee2018deep,Santos:20}, use spatially varying pixel exposures which trade spatial resolution for dynamic range~\cite{nayar2000high,nayar2003adaptive,nayar2006programmable,wetzstein2010sensor,hajisharif2015adaptive,serrano2016convolutional,alghamdi2019reconfigurable,Martel:2020:NeuralSensors}, or use optical encoding approaches that blur the sensor image~\cite{Rouf:2011,metzler2020deep,sun2020learning}. Specialized sensors, for example recording logarithmic irradiance~\cite{loose2001self} or floating point extended dynamic range values~\cite{Yang1999A6C} have also been proposed, but these either trade extended dynamic range for precision or require additional bandwidth.

\begin{figure}[t]
\begin{center}
  \includegraphics[width=\columnwidth]{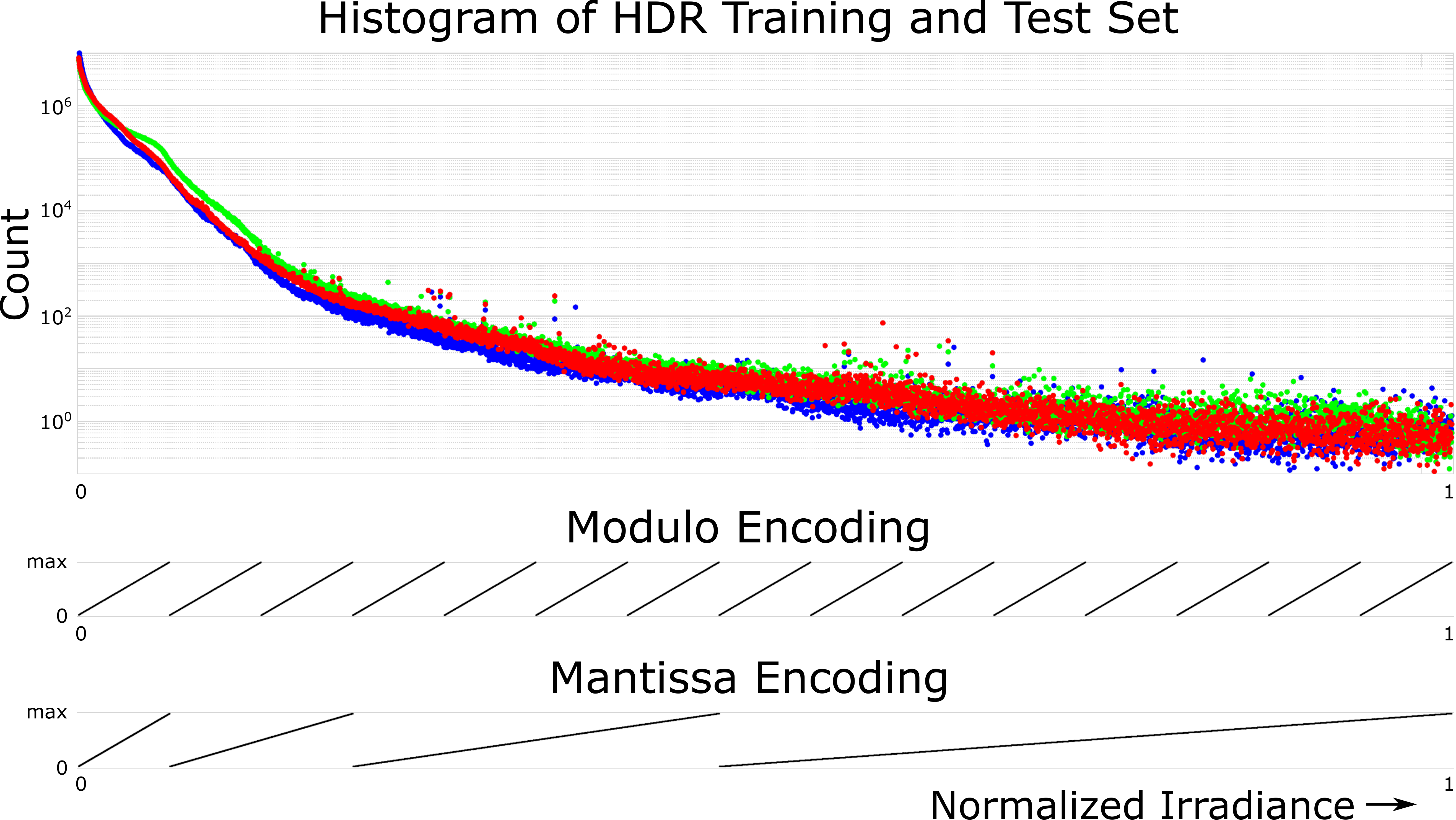}
\end{center}
  \caption{Log histogram of normalized irradiance values of all pixels in our training and test sets of HDR images for all color channels (top). This histogram is highly biased towards low-intensity values, indicating that irradiance values of natural images are not uniformly distributed. Yet, the modulo encoding subdivides this intensity range uniformly and wraps each of these areas into the available dynamic range of the sensor, as shown for a 1D ramp (center). The proposed mantissa encoding wraps the same 1D ramp in a perceptually more uniform manner in log space, which is observed as non-uniform wrapping in irradiance space (bottom). }
\label{fig:histogram}
\end{figure}

Our work (Fig.~\ref{fig:teaser}) is inspired by the idea of electronically applying a modulo encoding of the irradiance on the sensor followed by an intensity unwrapping algorithm~\cite{Zhao:2015,zhou2020unmodnet}. This idea is beneficial over other snapshot approaches, because it does not degrade a low-dynamic-range (LDR) image, as optical encoding approaches do, it does not hallucinate detail but recovers them, it does not decrease image resolution, or increase the required bandwidth. As we show in this paper, there are several downsides to the modulo camera, as proposed in prior work. Specifically, modulo wrapping is done directly in irradiance space, which allocates precision and number of wraps linearly in this domain. However, the human visual system is perceptually approximately linear in the log-domain, so a conventional modulo encoding wastes precision for detail that we do not perceive. Moreover, the irradiance distribution of natural scenes is heavily skewed towards darker values (see log-histograms in Fig.~\ref{fig:histogram}), so it makes sense to nonlinearly distribute the irradiance wraps in order to minimize their number, because they have to be computationally unwrapped again. 

\begin{figure}[t]
\begin{center}
  \includegraphics[width=\columnwidth]{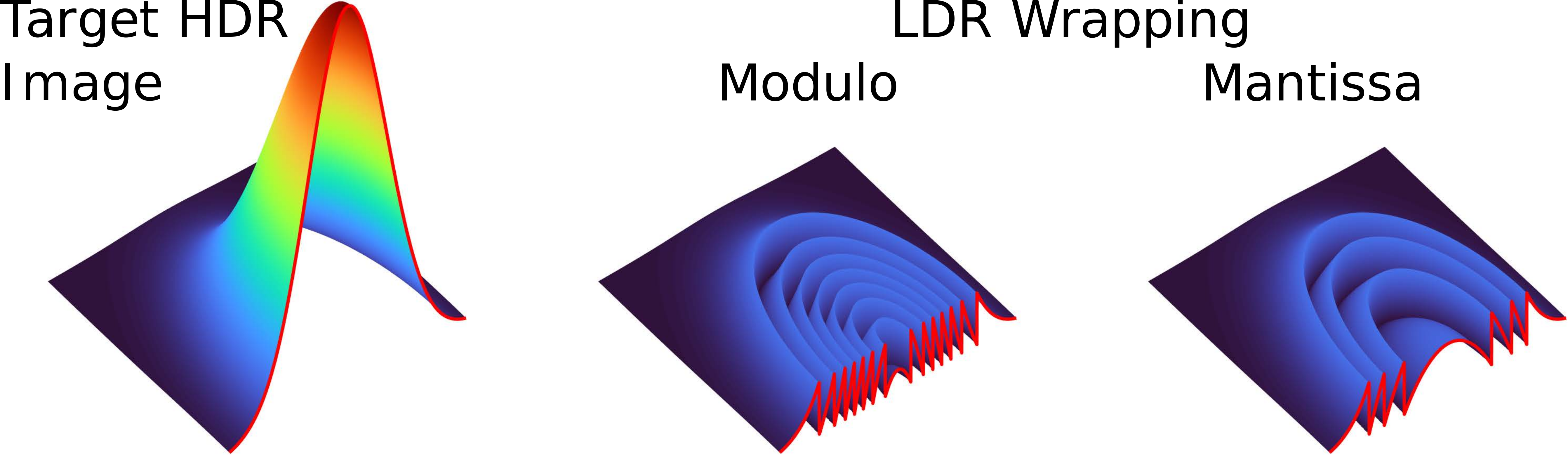}
\end{center}
  \caption{Example showing an HDR Gaussian function wrapped using the modulo and mantissa encoding in an LDR image. For this example, the modulo encoding requires more wraps than the mantissa encoding, which makes its reconstruction via computational unwrapping more challenging.}
\label{fig:3Dwraps}
\end{figure}

We address these challenges by proposing a perceptually inspired modulo-type wrapping scheme that operates in the log-irradiance domain. This idea intuitively combines the principles of operation of both log~\cite{loose2001self} and modulo~\cite{Zhao:2015} cameras. Indeed, the signal we propose to measure is essentially a generalization of the mantissa used by the  IEEE Standard for Floating-Point Arithmetic~\cite{ieee:2019}, or the log base 2 of the intensity modulo the well capacity. We demonstrate that such a log-modulo or \emph{mantissa} camera allocates precision in a perceptually meaningful manner and it nonlinearly distributes the wraps in irradiance space to better match the distribution of irradiance values in natural scenes (see Fig.~\ref{fig:histogram}, top). This directly leads to fewer wraps of natural scenes (see Figs.~\ref{fig:histogram}, center and bottom, and~\ref{fig:3Dwraps}), which make the inverse problem of 2D irradiance unwrapping easier to solve. To solve the unwrapping problem, we introduce a neural network architecture that is more robust than prior work using graph cut algorithms~\cite{Zhao:2015} or other network architectures~\cite{zhou2020unmodnet}. 
Finally, we prototype a {modulo} camera using a SCAMP-5 programmable sensor~\cite{carey2013100} which allows flexible re-configuration of the in-pixel irradiance encoding in software. These types of programmable sensors are expected to be widely available in the near future.


Specifically, we make the following contributions
\begin{itemize}
  \item We introduce MantissaCam as a new snapshot approach to HDR imaging, combining perceptually motivated irradiance encoding and decoding.
  \item We develop a neural network architecture that outperforms existing unwrapping methods for modulo cameras and that demonstrates state-of-the-art performance with our mantissa encoding. 
  \item We build a prototype {modulo camera and show improved results over previous methods.}
\end{itemize}

\noindent \textbf{Overview of Limitations.}

The SCAMP sensor we have does not include the log circuitry needed for capturing mantissa images, but we still demonstrate the benefits of the proposed reconstruction algorithm on captured modulo images. 

%% file: latex/sections/related.tex
\noindent \textbf{HDR Imaging.} 
The limited dynamic range of conventional camera sensors has been addressed by a number of computational imaging techniques. Exposure bracketing, for example, fuses several low-dynamic-range (LDR) photographs into a single HDR image~\cite{Picard95onbeing,Debevec:1997,mertens2009exposure,Hasinoff:2010,gupta:2013,HDRplus}. Temporally varying exposures can also be processed to obtain HDR videos~\cite{Kang:2003,Sen:2012:RPH,HDRVideo}. Yet, slight movements in the scene will create ghosting artifacts, which are challenging to be removed~\cite{tursun2015state}. Another class of approaches involves multiple sensors to capture these LDR images simultaneously~\cite{DBLP:journals/ijcv/AggarwalA04,McGuire:2007,Tocci:2011}. Although successful, these systems are expensive, bulky, and often difficult to calibrate.

Several approaches have been developed to estimate an HDR image from a single input image. Reverse tone mapping approaches aim at inverting a tone mapping operator~\cite{Banterle:2006,Meylan:85838,Rempel:2007}, which is an ill-posed inverse problem. Convolutional neural networks can also be directly applied to an LDR image to hallucinate the HDR image~\cite{marnerides2018expandnet,HDRCNN,Endo:2017,lee2018deep,Santos:20}. Neither of these approaches, however, has the capability to recover true image details. Bright highlights can also be optically encoded in an LDR image~\cite{Rouf:2011,metzler2020deep,sun2020learning}, but this approach relies on the required deconvolution to clean up even an LDR scene perfectly to compete with the quality of conventional sensors, which is challenging. Spatially varying pixel exposures are a promising direction but, similar to color filter arrays, they trade spatial resolution for dynamic range~\cite{nayar2000high,nayar2003adaptive,nayar2006programmable,wetzstein2010sensor,hajisharif2015adaptive,serrano2016convolutional,alghamdi2019reconfigurable,Martel:2020:NeuralSensors}.

Among these, our approach to snapshot HDR imaging is most closely related to the modulo camera~\cite{Zhao:2015}, which combines a modulo-type encoding of the irradiance on the sensor combined with a reconstruction algorithm that solves a 2D unwrapping problem. A conventional modulo operation, however, makes it difficult to distinguish between wrapping boundaries and high-frequency image detail. We introduce a perceptually motivated intensity wrapping technique for this class of computational cameras, which better preserves high-frequency image detail and dynamic range, and we also improve upon existing 2D upwrapping algorithms developed for related tasks.

\noindent \textbf{Unwrapping Algorithms.}  
Phase unwrapping is a problem often encountered in optical interferrometry, where the surface profile of some optical element or scene can be indirectly imaged as the wrapped phase of a coherent reference beam. A number of algorithms to unwrap these interferrograms has been developed, as surveyed in~\cite{ghiglia:98}.
When working with wrapped intensities of natural images, instead of optical phase values, the complex interplay of high spatial frequencies and drastically varying light intensity has to be accounted for. Unwrapping techniques for natural images has been analyzed~\cite{bhandari2020unlimited} and tailored algorithms developed~\cite{lang2017robust,shah2019signal,shah2017reconstruction}, but these require multiple input images. Most recently, the UnModNet network architecture was introduced to unwrap a single intensity image with state-of-the-art quality~\cite{zhou2020unmodnet}. Our network architecture improves upon this method for HDR imaging for modulo cameras but shows best results when used with the proposed mantissa encoding scheme.

\noindent{\textbf{Floating point sensors} from the early 2000s allow for capturing high dynamic range with multiple sampling \cite{ASerafinimultiple}, \cite{yang1999640} and variations with overlapping integration intervals\cite{ASerafiniPredmultiple}, or choosing optimum integration time \cite{Rhee}. Floating point sensors have great potential, however they require additional bandwidth. Our work reconstructs an HDR image from a captured image of the same bit depth as a conventional LDR sensor, utilizing the programmability of new sensors for in-pixel irradiance encoding together computational post-processing of that data. }

\noindent \textbf{Exotic Sensors for HDR Imaging.}
Specialized sensor circuits have been developed to support spatially varying pixel and adaptive exposures~\cite{mase2005wide,yang1999640,dudek2006adaptive,wagner2004high,martel2016parallel} as well as logarithmic~\cite{loose2001self} or modulo~\cite{Tyrrell:2009,Brown:2010,Zhao:2015} irradiance encoding. 
%
%
Emerging photon-counting sensors can facilitate HDR imaging but require high-speed readout circuitry and are best suited for low-light applications~\cite{Fossum:2016} or observe response functions that are similar to logarithmic sensors~\cite{ingle2019high}. 
All of these systems are inflexible, because they are not programmable. Near-focal-plane sensor--processors~\cite{zarandy2011focal} include some amount of computing capabilities in the sensor and related systems have become programmable~\cite{miao2008programmable,lopich201080,rodriguez2010cmos,zhang2011programmable,fernandez2012flip,carey2013100}. In this work, we use one of these platforms, SCAMP-5~\cite{carey2013100}, to prototype modulo encoding and the proposed neural network--based HDR reconstruction algorithm experimentally.

%% file: latex/sections/method.tex
\begin{figure*}[t]
\begin{center}
   \includegraphics[width=\linewidth]{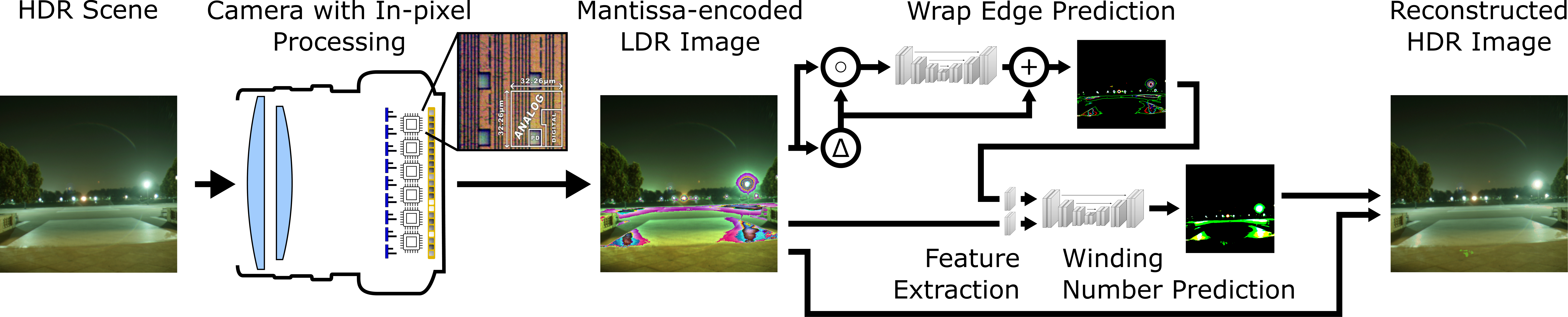}
\end{center}
   \caption{MantissaCam pipeline. An HDR scene is imaged by a camera with in-pixel processing capabilities, implementing the proposed irradiance encoding scheme (left). The resulting LDR sensor image encodes lower irradiance values similar to a conventional camera, but bright image regions, including the lamp and the reflections on the ground, are wrapped rather than saturated (center). The mantissa-encoded image is first processed by a network that predicts the wrap edges and then by another network that predicts the winding number (center right). The per-pixel winding number, together with the mantissa-encoded image, are used to reconstruct the HDR image (right). The symbols $\Delta$, $\circ$, and $+$ denote channel-wise Laplacian operators, channel concatenation, and addition, respectively.}
    \label{fig:pipeline}
\end{figure*}

The MantissaCam framework comprises an electronic in-pixel irradiance encoding scheme and a neural network--based decoding algorithm, which solves the 2D unwrapping problem to reconstruct the irradiance incident on the sensor. We discuss these aspects next.


\subsection{In-pixel Irradiance Encoding}

The image formation model of the MantissaCam is 
\begin{equation}
	I_{\textrm{sensor}} \left(x,y \right) = q \left( \textrm{mod} \left( \textrm{log}_\alpha \left( I \left(x,y \right) \right), I_{\textrm{max}} \right) \right)+ \eta ,
\end{equation}
where $I$ describes the spatially varying irradiance (i.e., the target HDR image) on the sensor, $I_{\textrm{sensor}}$ is the measured LDR sensor image, and $\eta$ is zero-mean additive Gaussian noise. The parameter $\alpha$ models a family of logarithmic irradiance response functions. For example, the special case $\alpha=2$ of our encoding scheme is similar to the mantissa encoding of the IEEE 754 standard for floating point arithmetic. Sensor quantization is modeled by the function $q( \cdot )$. $I_{\textrm{max}}$ is the maximum allowed irradiance value before the intensity wraps. This could be the well capacity of a pixel or a user-defined value that is slightly lower than that.

\subsection{Irradiance Decoding}

The proposed decoding scheme is implemented by two neural networks. The first takes the wrapped sensor image as input and predicts the wrap edges, effectively separating them from the texture edges. The second network predicts the winding number (i.e., the number of times intensity has wrapped) of each pixel from these wrap edges. 

To predict either modulo or mantissa wrap edges from a sensor image, we directly use the ``modulo edge separator'' proposed as part of the UnModNet architecture~\cite{zhou2020unmodnet}. This edge separator is a residual-type convolutional neural network (CNN) that takes as input a concatenation of the LDR sensor image and a Laplacian-filtered copy of the same. We illustrate our network in Figure~\ref{fig:pipeline} and refer the interested reader to~\cite{zhou2020unmodnet} for additional details.

Our second network predicts the winding number for each pixel, $W(x,y)$, given the wrap edges and the sensor image as input. For this purpose, features are extracted from both input images using the lightweight CNN-based feature extraction layers from~\cite{zhou2020unmodnet}. These are fed into an attention UNet~\cite{oktay2018attention} with four downsampling and four upsampling blocks, with each downsampling block using a strided convolutional layer and a residual bottleneck block, and each upsampling block mirroring it but with the addition of attention gates. This is a standard neural network architecture, but its application to directly predicting the winding number of irradiance-wrapped images is new. Note that this part of our algorithm is substantially different from the iterative, graph-cuts inspired unwrapping procedure proposed in~\cite{zhou2020unmodnet}. Their method aims at unwrapping the HDR image layer by layer, which is prone to propagating errors, whereas our approach directly predicts the number of wraps, i.e., the winding number, using a single pass through the UNet. We discuss additional details of this network architecture in the supplement and outline the training procedure of both networks in Section~\ref{sec:methods:details} and the supplement.

Given the predicted winding number for each pixel as well as the raw sensor $I_{\textrm{sensor}}$, we formulate the reconstruction of the HDR image $\widetilde{I}$ as
\begin{equation}
    \widetilde{I} \left(x,y \right) = \alpha^{ I_{\textrm{sensor}} + W(x,y) \cdot I_{\textrm{max}}}.
\end{equation}
%
In our implementation, we choose $\alpha = 2$. 

\subsection{Understanding the Relation between Resolution and Dynamic Range}

The theory addressing the ability to perfectly reconstruct a signal with MantissaCam falls within the framework of unlimited sampling recently developed in \cite{bhandari2017unlimited,bhandari2020unlimited}. Here, rather than formally treating the reconstruction problem, we attempt to highlight the advantages of a mantissa over a modulo encoding and develop an understanding of the tradeoffs between those.
\newcommand{\iimax}{I_{\mathrm{max}}}
\newcommand{\isense}{I_{\mathrm{sensor}}}
\newcommand{\modu}{\mathrm{mod}}
\newcommand{\mant}{\mathrm{mant}}
\newcommand{\wmod}{\mathcal{W}_\modu}
\newcommand{\wmant}{\mathcal{W}_\mant}
\newcommand{\wrap}{\mathcal{W}}
\newcommand{\dd}{\mathrm{d}}

Let us consider the 1D band-limited irradiance function $I(x)$, with maximal frequency $f_{\mathrm{max}}$. The irradiance is encoded on the sensor by the wrapping function $\wrap$ of the imaging model:
\begin{equation}
    \wrap:I\in\mathbb{R}_+ \mapsto \mathcal{W}(I) \in [0,\iimax].
    \label{eq:modulo}
\end{equation}
In particular, we consider the two wrapping functions:
\begin{equation}
    \wmod(I) = I - W\big(I(x)\big)\cdot \iimax, 
    \label{eq:mantissa}
\end{equation}
and
{\begin{equation}
  \wmant(I) = \log_\alpha(I) - W\big(\log_{\alpha}I(x)\big)\cdot \iimax,
\end{equation}
with $W\big(\cdot\big)=\left\lfloor \frac{\cdot}{\iimax} \right\rfloor $ and $\lfloor \cdot \rfloor$ being the floor function.}

In order to avoid aliasing on our discrete sensor array, we assume the sampling of $I$ respects the Nyquist sampling criterion $f_s>2\cdot f_{\mathrm{max}}$, with the sampling frequency $f_s$ related to the inverse pixel pitch $T_s=\frac{1}{f_s}$ (i.e. the resolution or pixel density, for instance expressed in line pairs per millimeter) of the sensor array.

\noindent \textbf{Recoverability of irradiance from modulo and mantissa encodings}
To get an intuition about the irradiance fields $\wrap(I)$ that can be perfectly reconstructed, let us consider the discretized irradiance $I[n]=I(n\cdot T_s)$ as seen by a pixel $n$. 

If a wrap of $\wrap(I)$ occurs within a pixel, information is lost and it is impossible to reconstruct the incident irradiance field. Therefore, a set of conditions to recover the field is:
\begin{equation}
\begin{cases}
    | \wrap(I[n+1])-\wrap(I[n]) | \leq \iimax, \\
    {|W(I[n+1])-W(I[n])| \leq 1,}
\end{cases}
\end{equation}
where the first condition derives from the Euclidean Division Theorem and makes sure we cannot wrap ``within'' a pixel, the second condition allows at most one wrap between two pixels. 

For the modulo encoding those conditions translate into 
\begin{equation}
   \big| I[n+1] - I[n] \big| \leq \iimax,
\end{equation}
and for the mantissa encoding we have that
\begin{equation}
   \big| \log_\alpha(I[n+1]) - \log_\alpha(I[n]) \big| \leq \iimax.
\end{equation}

This shows that while the modulo encoding can reconstruct any irradiance with arithmetic growth of $\iimax$, a mantissa encoding can reconstruct a larger class of functions with geometric growth of $\iimax$. 

\noindent \textbf{Dynamic range.} For both types of encoding, these results imply an interesting tradeoff between the dynamic range of the sensor and its spatial resolution. With two sensors of the same size, using different pixel pitches $T_s$ and $T_s'$ such that $T_s'>T_s$, the sensor with a smaller pixel pitch $T_s$ (i.e., of higher resolution) can reconstruct faster spatial variations of irradiance ($\frac{\iimax}{T_s}>\frac{\iimax}{T_s'}$ in the modulo case). Therefore, there is a relationship between the maximum dynamic range recoverable for a sensor given its resolution. For two sensors of fixed size with $N$ pixels, the maximum recoverable irradiance is a ramp starting at pixel $n=0$ and ending at pixel $n=N-1$. In this setting, the sensor with modulo encoding can reconstruct a maximum dynamic range of $\mathrm{DR}\approx 10\log(N\cdot\iimax)$dB while the one with a mantissa encoding can recover a much wider dynamic range of $\mathrm{DR}\approx 10\cdot N\log(\iimax)$dB.

\noindent \textbf{Quantization.}
The ultra-high dynamic range of the mantissa encoding comes at the expense of loss of precision. In practice, no sensor has infinite bit depth but is quantized to 8--12~bits. As shown in the bottom graphs of Figure~\ref{fig:histogram}, the same number of levels are distibuted on a much wider range as the winding number 
$W$ increases. This means a MantissaCam cannot resolve irradiance with the same precision ModuloCam can at high irradiance levels---the quantization error is higher for our encoding. Yet, early psycho-physics studies~\cite{fechner1948elements} noted that perceived light intensity is proportional to the logarithm of the light intensity. Known as Fechner-Weber law, this implies that the coarser quantization of MantissaCam at high irradiance levels might not be perceptually important.

\begin{figure*}[t]
\begin{center}
   \includegraphics[clip, trim=0cm 10cm 0cm 0cm, width=1\linewidth]{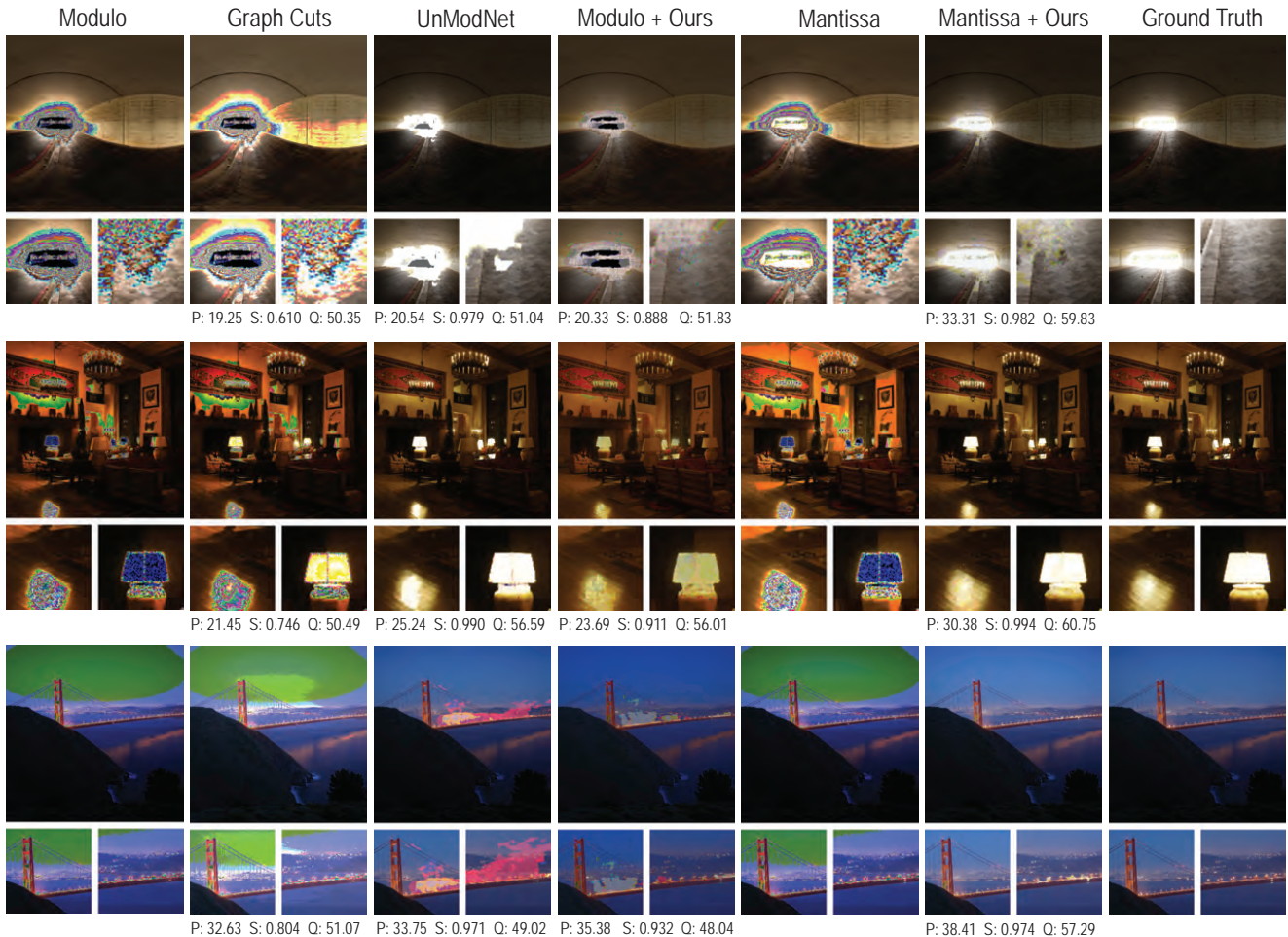}
\end{center}
   \caption{Evaluation of encoding and decoding schemes in simulation. A conventional modulo encoding wraps the irradiance of a scene into an LDR sensor image (column~1). A graph cuts--based reconstruction algorithm~\cite{Zhao:2015} usually performs poorly (column~2) whereas the recently proposed UnModNet architecture~\cite{zhou2020unmodnet} often estimates reasonable HDR images (column~3). Yet, the proposed reconstruction framework works best among these methods (column~4). Moreover, the proposed mantissa encoding scheme (column~5) induces fewer irradiance wraps making it easier to reconstruct the HDR image using our framework (column~6). Our approach achieves reconstructions closest to the ground truth (column~7). `P', `S', and `Q' indicate the PSNR, SSIM and Q-score for each reconstruction method.
   }
\label{fig:results}
\end{figure*}

\subsection{Dataset and Implementation Details}
\label{sec:methods:details}


For a fair comparison, the dataset used to train and evaluate our model was the same dataset created by UnModNet~\cite{zhou2020unmodnet}. We randomly split the images into 400 training images and 193 testing images. We used the same process to augment the training dataset, over-exposing and cropping images to yield a total of 5,945 training images. 

We train our networks in three stages. First, we train the wrap edge prediction network by itself for 400 epochs, taking simulated sensor images as input, using a binary cross entropy loss with the ground truth wrap edge images obtained via simulation. Second, we train the winding number prediction network by itself for 200 epochs, taking simulated sensor images and ground truth wrap edges as input, using a mean-squared error (MSE) loss on the ground truth winding number. Third, we train both networks end-to-end for another 200 epochs using the same MSE loss on ground truth winding number. Additional implementation details are found in the supplement.



%% file: latex/sections/experiments.tex
\begin{table}[b]
\footnotesize
\begin{center}
\begin{tabular}{@{\hskip -1pt}l@{\hskip -1pt} c c c c c@{\hskip -0.1pt} c} 
 \toprule
	Encoder & \multicolumn{3}{c}{Modulo}  & Mantissa $\!\!\!$ & $\!\!\!$ None \\ 
  Decoder & Graph Cuts~\cite{Zhao:2015} $\!\!\!\!$ & $\!\!\!\!$ UnModNet~\cite{zhou2020unmodnet} $\!\!\!\!\!\!\!\!$ & $\!\!\!\!$ Ours $\!\!\!\!\!\!\!\!$ & $\!\!\!\!\!\!\!\!$ Ours $\!\!\!\!\!\!\!\!$ & $\!\!\!\!\!\!\!\!$ CNN~\cite{HDRCNN} \\ 
	\midrule 
 PSNR ($\uparrow$) &  21.4 & 29.5 & 32.2 & \bf{37.4} & 22.7* \\  
 Q Score ($\uparrow$) &  48.0 & 59.1 & 57.1 & \bf{60.9} & 47.7*  \\
  SSIM ($\uparrow$) &  0.80 & 0.79 & 0.84 & \bf{0.97}   & 0.72* \\
  MSSIM ($\uparrow$)  & 0.82 & 0.91 & 0.93 & \bf{0.99}   & 0.76*\\
 LPIPS ($\downarrow$) &  0.29 & 0.12 & 0.10 & \bf{0.03}  & ---  \\ 
 \bottomrule
\end{tabular}
\end{center}
\caption{Quantitative evaluation of modulo and mantissa in-pixel encoding combined with various reconstruction algorithms for simulated data. Our irradiance unwrapping network performs better than existing algorithms on the modulo encoding, as evaluated by several metrics. Combined with the proposed mantissa encoding, our approach achieves state-of-the-art results. We also show the quality of a CNN working with conventional LDR images using the same dataset. Values marked with * are reproduced from~\cite{zhou2020unmodnet}.}
\label{tab:quantitative}
\end{table}

\subsection{Evaluation on Synthetic Data}

Figure~\ref{fig:results} qualitatively and quantitatively compares modulo and mantissa encoding schemes combined with different reconstruction algorithms. Using a single modulo-wrapped image as input, graph cuts perform poorly~\cite{Zhao:2015}. The UnModNet network~\cite{zhou2020unmodnet} does reasonably well in some cases, but struggles to reconstruct the large bright parts of the first example scene and the lights on the bridge of the third scene. {Their iterative unwrapping procedure sometimes fails in stopping to unwrap, which results in extremely high irradiance values lowering their PSNR and obscuring fine image detail.} Our algorithm achieves a better quality than these methods on the same modulo-encoded images, as evaluated by the peak signal-to-noise (PSNR or P), structural similarity (SSIM or S), and Q-score of the perceptual HDR-VDP-2~\cite{Mantiuk:2011} metrics. Moreover, when combined with the proposed mantissa irradiance encoding scheme, our framework achieves the best results among all of these methods.

Table~\ref{tab:quantitative} also quantitatively compares all of these approaches using several different metrics on the test set of the dataset described in Sec.~\ref{sec:methods:details}. In addition to the above methods, we also include a comparison to a CNN operating directly on a conventional LDR sensor image to hallucinate the HDR scene~\cite{HDRCNN}. {Not shown are the results from the combination of the UnModNet architecture with the mantissa encoding. The average PSNR was less than 10 dB due to UnModNet's iterative unwinding. It is prone to propagating errors and with the mantissa encoding, the errors are ``exponentially'' propagated. As shown in Table~\ref{tab:quantitative}}, the proposed mantissa encoding scheme combined with our reconstruction framework achieves the best results using all metrics, outperforming the state of the art, i.e., UnModNet, by almost $8$~dB of PSNR.

All simulations with synthetic data are run on noise-free images to study the upper bound of all of these algorithms. However, we do include results of simulations with simulated sensor noise in the supplement and also evaluate the best-performing algorithms on noisy captured data in the following.

\subsection{Prototyping a Modulo Camera using SCAMP-5}

We build a physical prototype using an example of an emerging class of sensors, dubbed focal-plane sensor--processors~\cite{zarandy2011focal}, that embed small processing circuits inside each pixel. We use SCAMP-5~\cite{carey2013100}, whose processing elements (PE) are programmable in a single instruction multiple data (SIMD) fashion, similar to a GPU where the same instruction is performed for all processing elements simultaneously on some local piece of data. Specifically, a PE is equipped with a few analog and digital memories. Instructions can be performed as light is being collected by the pixel's photo-sensitive element, thus enabling to change the way integration is performed, as required for our implementation. {In other SCAMP versions, there is log circuitry that would allow us to take mantissa images, however, our version does not have this capability. We are still able to implement the modulo camera and show the benefits of our reconstruction method over previous state-of-the-art methods.}

\subsection{{Experimental Results}}


\begin{figure}[t]
\begin{center}
   \includegraphics[trim=0cm 0cm 0cm 0cm, width=0.9\linewidth]{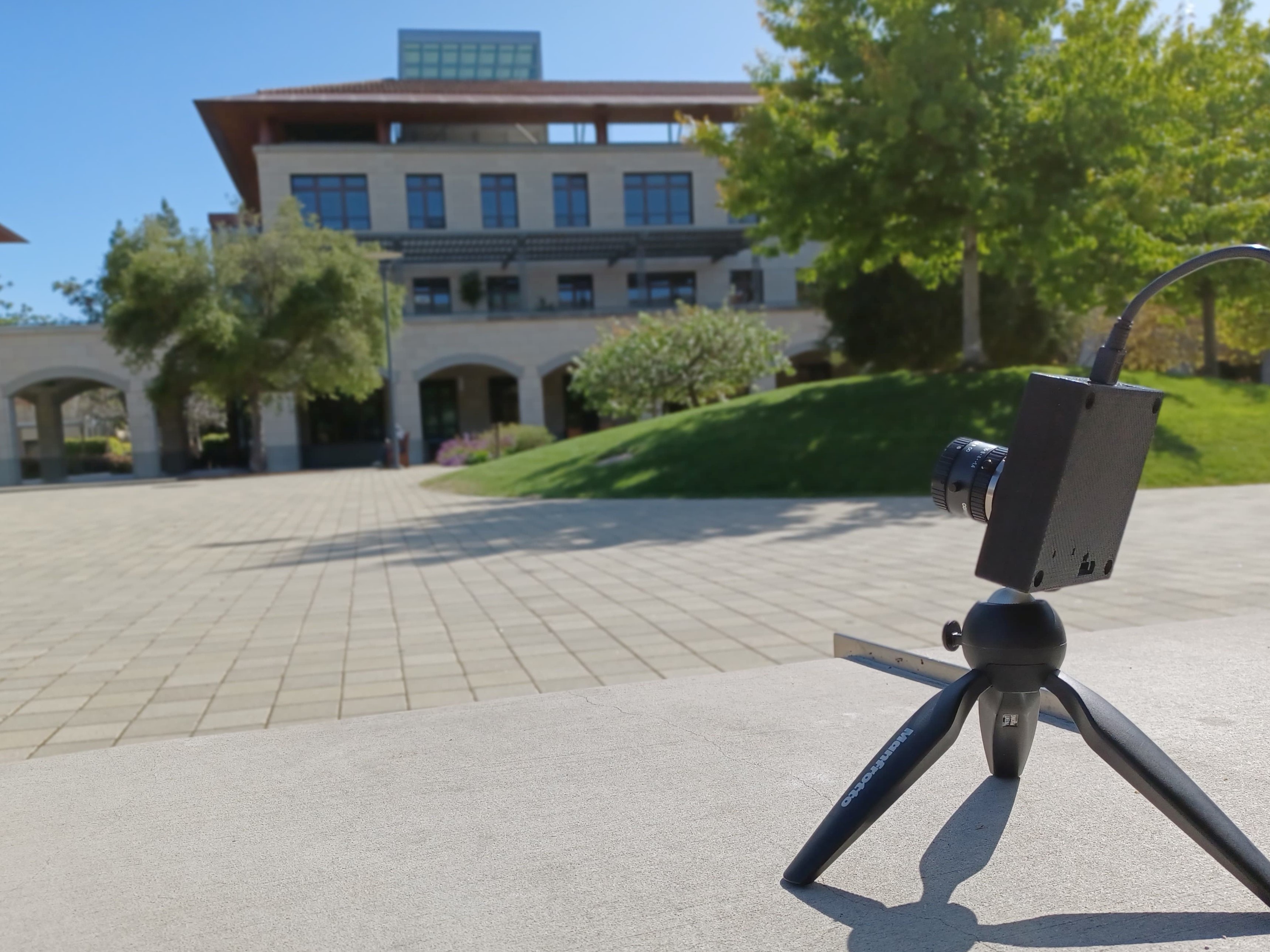}
\end{center}
   \caption{Prototype camera capturing an outdoor HDR scene.}
\label{fig:prototypephoto}
\end{figure}

We use SCAMP-5 to prototype {a modulo camera} and capture {HDR scenes outside} (see Fig.~\ref{fig:prototypephoto}). This sensor records grayscale images with a resolution of $256 \times 256$ pixels. For this experiment, {we retrained both UnModNet and our network on modulo images} using the same training procedure described in Section~\ref{sec:methods:details}, but on grayscale images captured with SCAMP-5. {For this purpose, we collected a dataset of 14,810 modulo and corresponding reference HDR images using the SCAMP-5 prototype. We split this dataset into 13,329 training images and 1,481 test images. No artificial data augmentation was performed. We trained a single edge predictor network that we used for UnModNet's iterative unwrapping approach and also as part of our own pipeline. This network was trained for 150 epochs using the experimentally captured dataset.}

\begin{figure*}[t]
\begin{center}
   \includegraphics[clip, trim=0cm 10.2cm 0cm 0cm, width=1\linewidth]{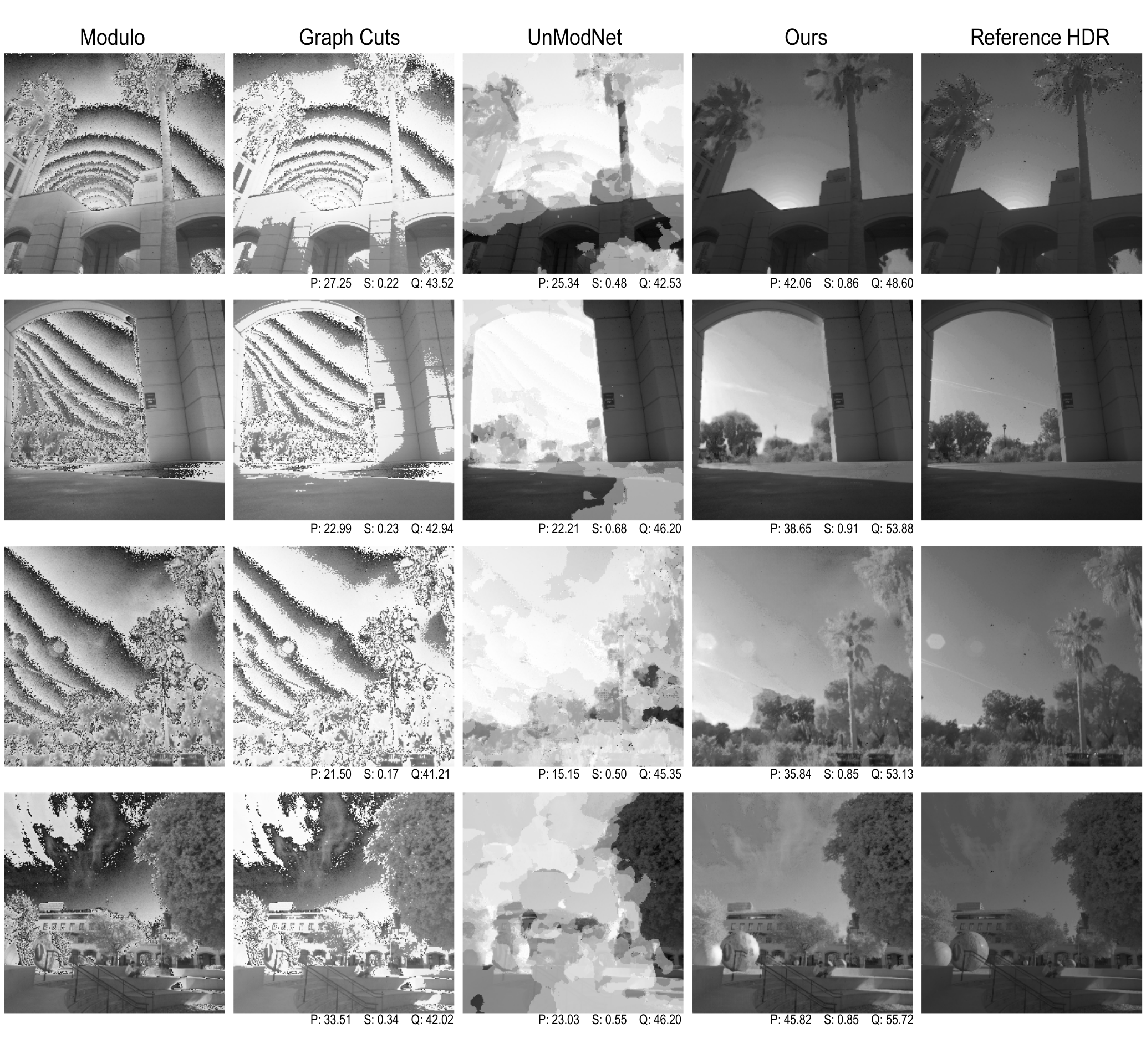}
\end{center}
   \vspace{-0.5cm}\caption{{Experimental results. Using a programmable sensor, SCAMP-5, we capture (noisy) modulo images (left) and process them using graph cuts, UnModNet, and our network applied to the captured modulo data. Tonemapped results using all these reconstruction methods as well as a reference HDR image are shown for several different scenes.}}
\label{fig:scampresults}
\end{figure*}

{Figure~\ref{fig:scampresults} shows captured modulo images, the tonemapped reconstructions, and a tonemapped reference HDR image. The captured images include sensor noise, which is especially noticeable around the irradiance wraps. The graph cuts and UnModNet algorithms usually fail to estimate reasonable HDR images, likely due to the noise in the sensor images. For more recognizable results, we limited the number of unwrappings for UnModNet to a maximum of five iterations. Otherwise, the reconstructions end up completely white. The dynamic range of this scene is far greater than that of the sensor, yet our method is able to reconstruct HDR images with high quality. }

{Table~\ref{tab:scamp_table} shows the comparison of graph cuts, UnModNet, and our method averaged over the test set captured with the SCAMP-5. We compare PSNR, Q score, SSIM, MSSIM, and LPIPS scores. Across all metrics, ours outperforms previous methods by a large margin. }

\begin{table}
\footnotesize
\begin{center}
\begin{tabular}{@{\hskip -1pt}l@{\hskip -1pt} c c c c c } 
 \toprule
	Encoder & \multicolumn{3}{c}{Modulo}  \\ 
  Decoder & Graph Cuts~\cite{Zhao:2015} $\!\!\!\!$ & $\!\!\!\!$ UnModNet~\cite{zhou2020unmodnet} $\!\!\!\!\!\!\!\!$ & $\!\!\!\!$  Ours $\!\!\!\!\!\!\!\!$ & $\!\!\!\!\!\!\!\!$ & $\!\!\!\!\!\!\!\!$ \\ 
	\midrule 
 PSNR ($\uparrow$) & 20.3  & 15.2 & \bf{33.7}  &   \\  
 Q Score ($\uparrow$) & 43.7  & 45.7 & \bf{53.0} &  \\
  SSIM ($\uparrow$) & 0.27  & 0.52 & \textbf{0.85} &     \\
  MSSIM ($\uparrow$)  & 0.23 & 0.59 & \textbf{0.95} &  \\
 LPIPS ($\downarrow$) & 0.14  & 0.12 & \bf{0.09} &  \\ 
 \bottomrule
\end{tabular}
\end{center}
\caption{{Quantitative evaluation of modulo in-pixel encoding combined with various reconstruction algorithms for experimentally captured data. Our algorithm processing the same modulo images as the others achieves significantly better results in all relevant metrics.}}
\label{tab:scamp_table}
\end{table}

\begin{figure*}[t]
\begin{center}
   \includegraphics[clip, trim=0cm 16cm 0cm 0cm, width=1\linewidth]{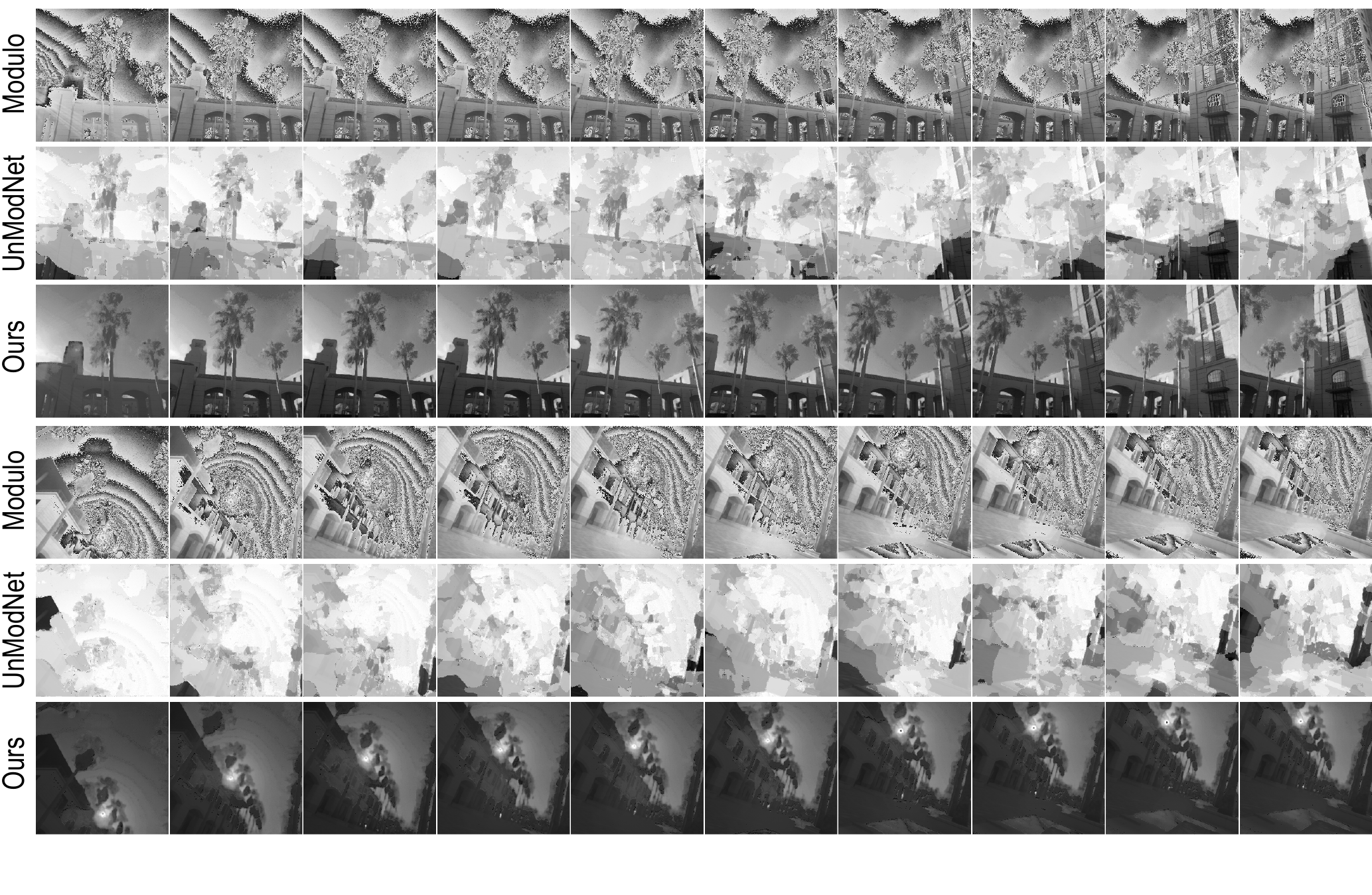}
\end{center}
   \vspace{-0.5cm}\caption{{HDR video experimental results. We show 10 frames of two captured modulo video sequences, UnModNet's reconstruction, and our reconstruction. Our reconstruction shows temporal consistency and good image quality whereas UnModNet typically fails to estimate reasonable results. }}
\label{fig:scampmotion}
\end{figure*}

{With our single-shot HDR image unwrapping method, we can also capture short HDR video clips, which would have been difficult to do with conventional HDR methods like bracketed exposures. In Figure~\ref{fig:scampmotion}, we show a sequence of modulo-encoded frames that we captured while moving the camera. We also show tonemapped reconstructions using UnModNet and our network. Our method unwraps the modulo video sequence with high temporal consistency and good quality, while lots of flickering and poor image quality are observed for UnModNet. Video clips of these and other example scenes are included in the supplemental material.}

%% file: latex/sections/discussion.tex
Motivated by the emerging class of programmable sensors, we demonstrate new capabilities they could enable for the long-standing challenge of snapshot HDR imaging. For this purpose, we develop a reconstruction algorithm for the modulo camera that is more robust and achieves better results than the current state of the art. Moreover, we introduce the mantissa encoding scheme that is inspired by the human visual system and achieves a favorable tradeoff between dynamic range, spatial frequency, and precision when encoding HDR scenes compared to the modulo camera. We evaluate our system in simulation but also show preliminary results captured with a prototype SCAMP-5 programmable sensor,  {demonstrating the effectiveness of our reconstruction algorithm on the modulo camera.} The global shutter speed in our simulations and with the prototype are always set to capture the desired level of detail in the dark regions, relying on the encoder and reconstruction algorithm to recover the brightest parts of the scene.

\begin{figure}
\centering
   \includegraphics[clip, trim=0cm 4cm 0cm 0cm, width=\columnwidth]{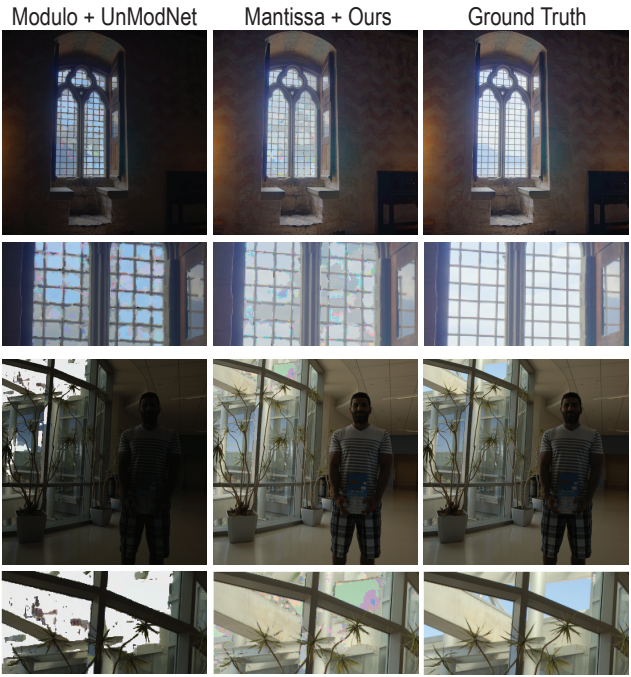}
   \caption{Limitations. Challenging areas for unwrapping often include regions with high spatial detail and wrapping or dense edges where it may be difficult for the networks to differentiate between wrap and texture edges. While our method is able to better reconstruct some of these areas than a modulo camera with the UnModNet algorithm, some artifacts remain.}
\label{fig:limitation}
\end{figure}

\noindent\textbf{Limitations and Future Work.}  Although promising, the proposed system has several limitations. First, our reconstruction pipeline improves results over existing work by a large margin, yet it fails in some cases as shown in Figure~\ref{fig:limitation}. Thus, there is room for further improving the robustness of the algorithm. Second, our mantissa-based encoding scheme is intuitive and robust, but the question of what an optimal encoding scheme for HDR imaging or other applications remains. Some prior work has studied end-to-end-optimized in-pixel irradiance encoding~\cite{Martel:2020:NeuralSensors}, which could be a fruitful direction for (un)wrapping-based HDR cameras, such as ours. Yet, optimizing periodic objective functions, such as modulo and mantissa-like functions, is not trivial and requires additional research. Third, the class of computational HDR cameras we discuss here seeks to improve the dynamic range of sensors for bright scene parts, but it does not necessarily improve the black level or performance in low-light conditions. It would be valuable to study how in-pixel intelligence offered by programmable sensors could help imaging in low-light scenarios, although this is beyond the scope of our work. {Fourth, in our experiments we ignore the effect of the color filter array (CFA), primarily because our prototype is grayscale.} 

Furthermore, our SCAMP-5 prototype has many hardware limitations, including a high read noise level, low pixel fill factor, low resolution, lack of color filters, and a challenging software interface. Improving these aspects with better circuit design, 3D fabrication techniques, and improved firmware engineering could make this or related platforms better and more accessible to the computational photography community. The programmable sensor is a valuable tool in early experimentation. Ultimately, it could be replaced by a specialized CMOS image sensor device, implementing, in hardware, the optimized version of the mantissa-like encoding.

\noindent\textbf{Conclusion.}
The emerging class of programmable sensors enables in-pixel intelligence, offering new imaging capabilities for computational photography systems. While our system demonstrates a new co-design of in-pixel irradiance encoding and decoding for snapshot HDR imaging, many other applications in computer vision, photography, and autonomous driving could be enabled by this platform. Our work takes first steps towards the vision of adaptive and domain-optimized computational cameras.

%% file: latex/supplement.tex
\newenvironment{Figure}
  {\par\medskip\noindent\minipage{\linewidth}}
  {\endminipage\par\medskip}

\onecolumn
\begin{center}
\section*{\Large Supplemental Material\\MantissaCam: Learning Snapshot High-dynamic-range Imaging with Perceptually-based In-pixel Irradiance Encoding}
\end{center}

\begin{multicols}{2}
\renewcommand{\thefigure}{S\arabic{figure}}
\renewcommand{\thesection}{S\arabic{section}}
\renewcommand{\thetable}{S\arabic{table}}

\section{Pipeline Details}

\subsection{Mantissa Dataset creation}
There are several ways to encode the mantissa. When working with synthetic data, the simplest way is to just take the $\log$ of the signal and then take the modulo of the resulting value. Recall $\log$ of anything below 1 is a negative value, which would not be conceivable with the hardware. Instead, only after the pixel saturates do we take the $\log$ to simulate the mantissa. When we saturate the pixel, the subsequent wrap will require twice the intensity to wrap again. To create the training dataset, we create the mantissa image and the corresponding winding number image. For each pixel $ij$,
\begin{equation}
        \text{mantissa}_{ij}= 
\begin{cases}
    I_{ij},& \text{if } I_{ij} < I_{max}\\
    \log_{\alpha}(I_{ij}) \%I_{max},& \text{otherwise.}
\end{cases}
\end{equation}
 
\begin{equation}
    \text{winding number}_{ij}= 
\begin{cases}
    0,&\text{if } I_{ij} < I_{max}\\
    \big\lfloor \log_{\alpha}(I_{ij}) \big\rfloor + 1,& \text{otherwise.} 
\end{cases}
\end{equation}
where $\%$ denotes the modulo operation and $\left\lfloor \cdot \right\rfloor$ denotes the floor function. For our dataset and experiments, we set  $I_{max} = 1$ and $\alpha = 2$.

\subsection{Network Architecture}
\begin{figure*}[t]
\begin{center}
   {\includegraphics[ width=\linewidth]{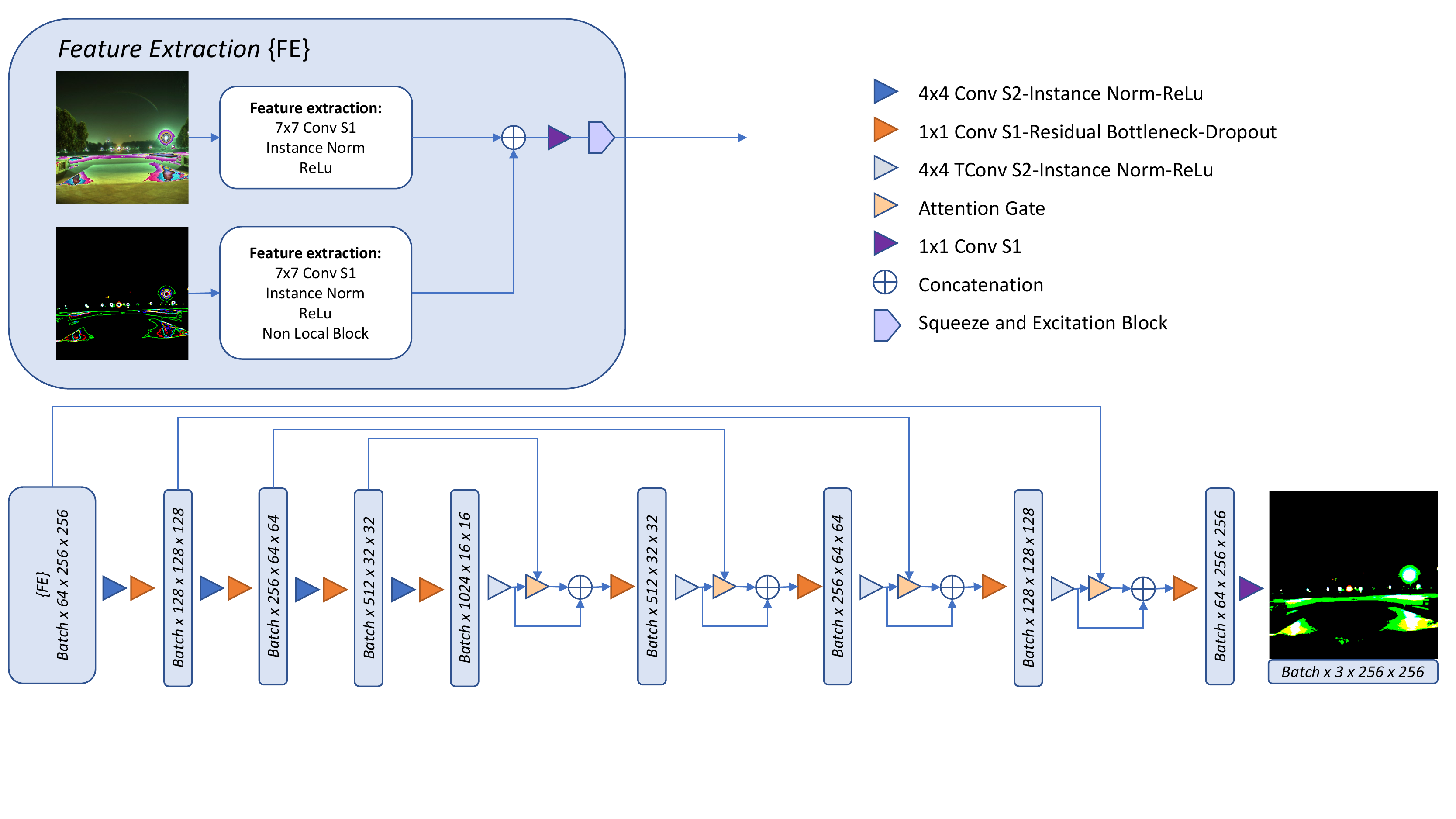}}
\end{center}
   \caption{Attention UNet architecture of the winding number prediction network. }
\label{fig:AttentionNetwork}
\end{figure*}

In this subsection, we describe the architecture for the single pass winding number prediction network (also see Figure~\ref{fig:AttentionNetwork}). The extracted edges from the edge prediction network, along with the mantissa image, are fed into the network via feature extraction by a $7 \times 7$ convolutional layer, an instance norm, ReLU, and a non-local block for the extracted edge features. These images are then concatenated and sent through a squeeze-and-excitation block to perform dynamic channel-wise feature recalibration. The base network is an attention unet, pioneered by Oktay et al.~\cite{oktay2018attention}. The backbone is the U-Net where the expanding path has attention gates added, along with the skip connections. Skip connections allow features extracted from the contracting path to be used in the expanding path. The attention block places more emphasis on the features of the skip connections. 

\section{Training Procedures}

\subsection{RGB training on HDR images}
For training our network for RGB, we trained the edge network for 400 epochs on the synthetic data at a learning rate of .0001 using an ADAM optimizer in Pytorch. From a dataset of 593 images, we randomly split it into 400 training images and 193 test images. We augment the training images by scaling the HDR image and calculating the corresponding mantissa and winding numbers.

\subsection{Training Procedure for SCAMP-5 Prototype}
We retrain the edge prediction network for the captured grayscale dataset as described in the paper. Both UnModNet and our method use the same edge prediction network. The other parts of the respective pipelines are retrained on the captured dataset using a similar procedure as used for the synthetic data described above.

\subsection{Baseline Comparison}

\paragraph*{Graph Cuts}
Graph Cuts was implemented following the original ModuloCam paper~\cite{Zhao:2015} using the same custom potential function. Reaching out to the authors confirmed the method, which can be successful for some clearly wrapped modulo images, however requires delicate parameter tuning for each of the many layer unwrappings of each image. We chose a set of parameters to best unwrap the whole set. PSNR, SSIM, and MSSIM scores were comparable to those found in UnModNet~\cite{zhou2020unmodnet} when they implemented the MRF algorithm.

\paragraph*{Modulo Encoding with UnModNet}
We retrained UnModNet, the state of the art for unwrapping modulo images, with the same training process and same dataset as in \cite{zhou2020unmodnet} and results were comparable to those reported in the paper. In areas of dense wrappings, the pipeline struggles to stop unwrapping, leading to patches of white. 

\paragraph*{Mantissa Encoding with UnModNet}
One of our baseline experiment is to use the pipeline of UnModNet with the forward imaging model of MantissaCam. We trained the pipeline using the same training procedure as described above. We noted the layer-by-layer unwinding did not work well with the reconstruction from the mantissa encoding as errors in winding number manifest in exponentially bad errors. Indeed, missing a wrap results in much worse errors in MantissaCam (because of the exponential function used when reconstructing) than in ModuloCam, resulting in huge artifacts. Besides, the nature of the layer-by-layer unwrapping is prone to propagating errors.

\paragraph*{Modulo Encoding with Our Network}
To combat propagation of errors in unwinding, we directly predict the winding number in a single pass through an attention-unet instead of predicting a mask. Again, we train using the same training procedure as UnModNet. Results are promising, however, the network still struggles when the modulo image has very tight wrappings (of the order of 1--2 pixels width). 

\paragraph*{Mantissa Encoding with Our Network}
Introducing the mantissa allows us to spatially spread out the wraps as we get to higher and higher irradiance levels. This leads to preservation of more detail. Results comparing these methods, excluding the UnModNet for the mantissa, are shown in the paper and in the additional figures in Section~4.

\subsection{Additional Implementation Details}
We compare the full reconstructed HDR image with the ground truth HDR image to calculate PSNR and Q-Score (2). We then tonemap both the ground truth and the predicted HDR images, all using the Reinhard Algorithm with gamma = 1, intensity = 1. The tonemapped images are then compared to calculate SSIM and MSSIM values. Inference time for our method is much faster than for the UnModNet or graph cuts due to the single pass architecture, as opposed to the iterative unwrapping that can unwrap as high as the default of 15 max iterations.

\begin{Figure}
  \includegraphics[trim=0cm .3cm 0cm 0cm,width=\linewidth]{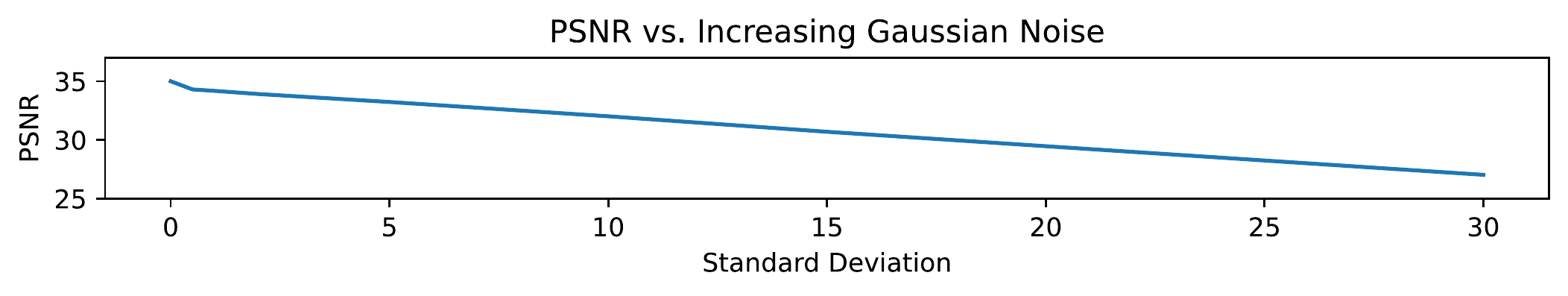}
  \captionof{figure}{my caption of the figure}
\label{fig:noise_study}
\end{Figure}

\section{Additional Details of Experimental Results}
Currently, mantissa images cannot be directly captured in SCAMP-5. However we implemented a procedure on SCAMP-5 to capture modulo images as described in the main paper. 

We also implemented a bracketed exposure procedure directly on the camera in order to get reference HDR images. Exposure bracketing is performed by capturing 5 images doubling the exposure time between each exposure, starting from a configurable short exposure time.

\section{Additional Results}
\begin{figure*}[t]
\begin{center}
   \includegraphics[width=0.95\linewidth]{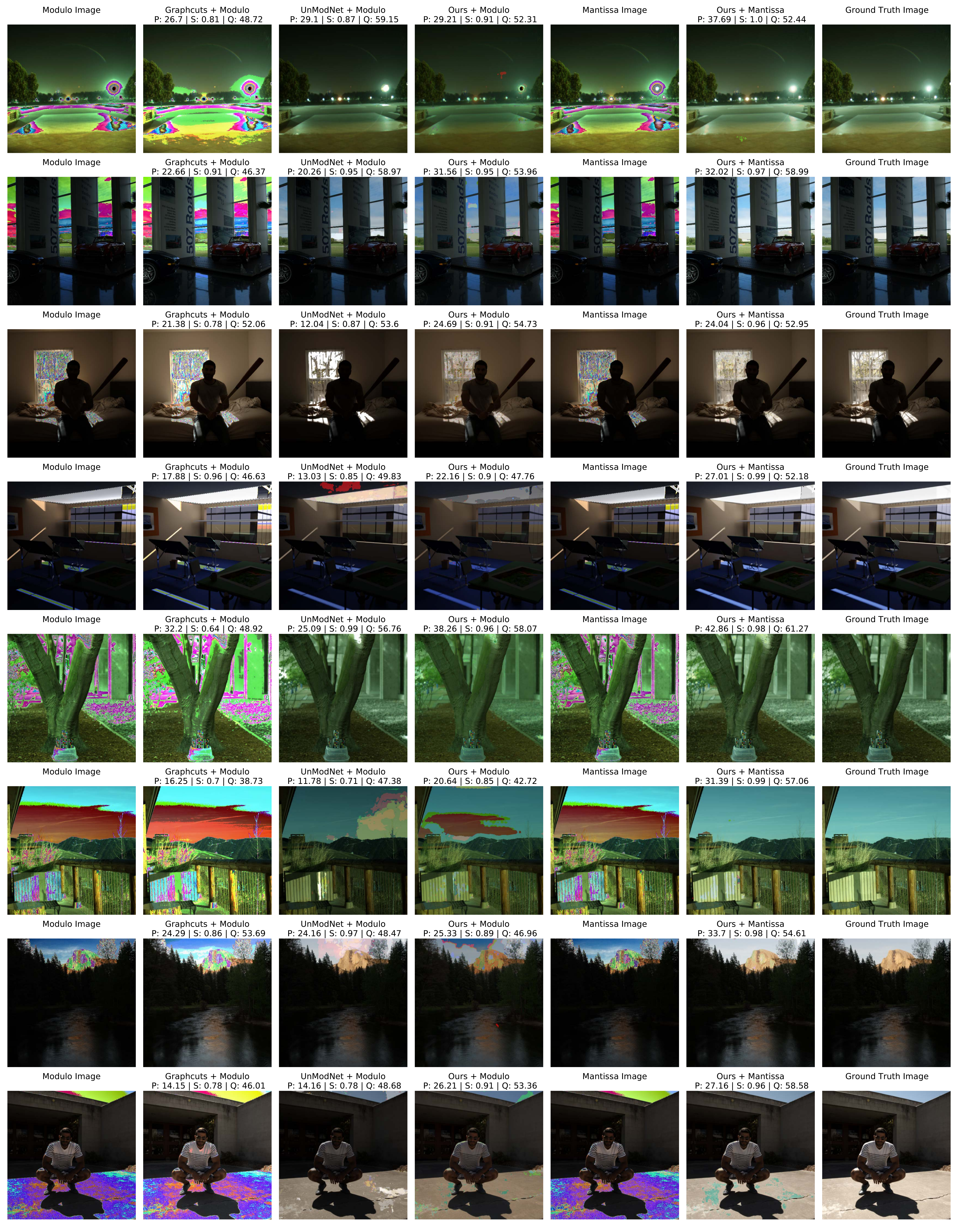}
\end{center}
   \caption{More results showing the comparison between different baselines and encodings. Ours + mantissa is better able to keep details in the high intensity areas. PSNR (P), SSIM (S), and Q-Scores (Q) are shown about the reconstructed images.}
\label{fig:supplement1}
\end{figure*}

\begin{figure*}[t]
\begin{center}
   \includegraphics[width=0.95\linewidth]{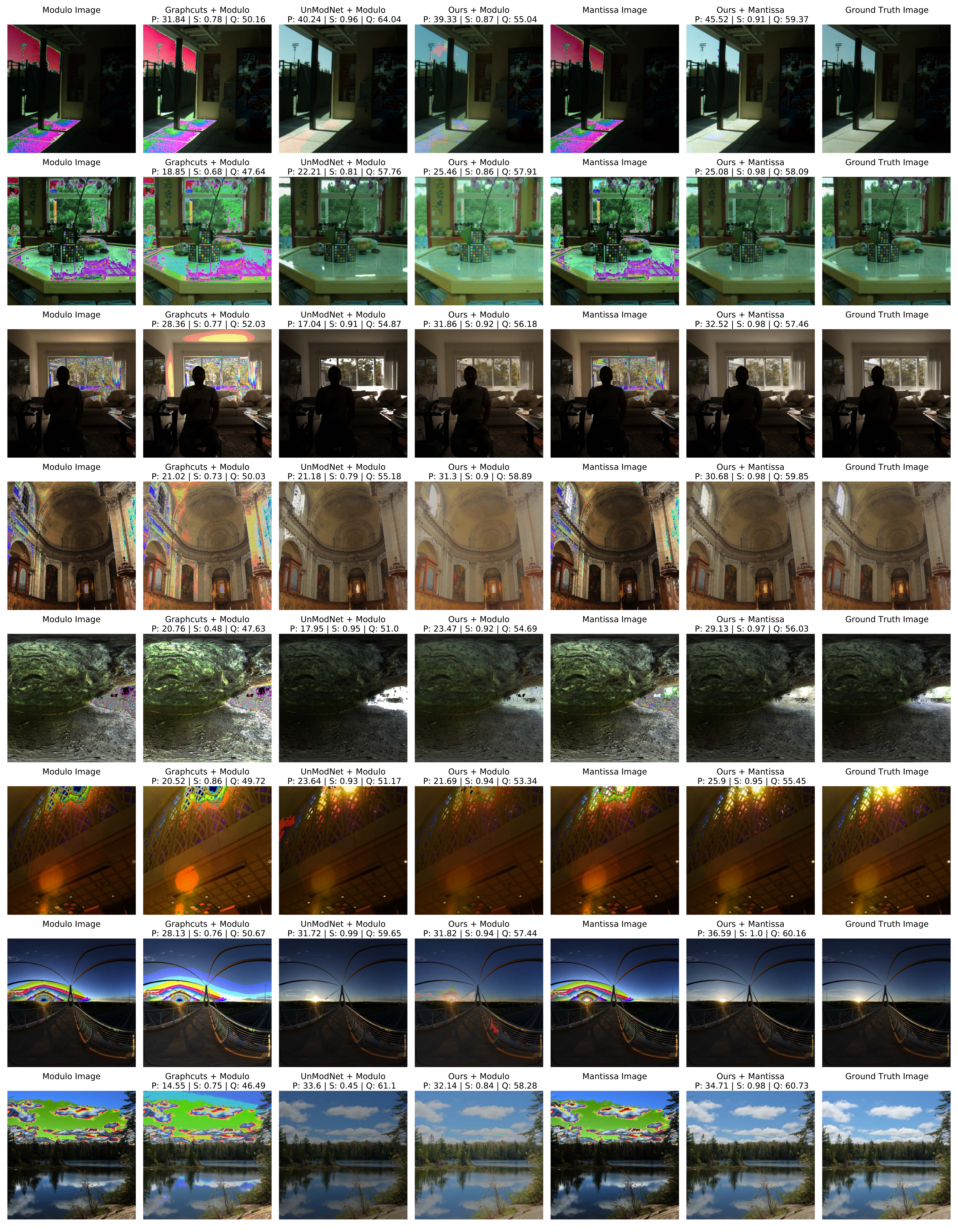}
\end{center}
   \caption{More results comparing the different reconstruction and encoding methods.}
\label{fig:supplement2}
\end{figure*}

\begin{figure*}[t]
\begin{center}
   \includegraphics[width=0.8\linewidth]{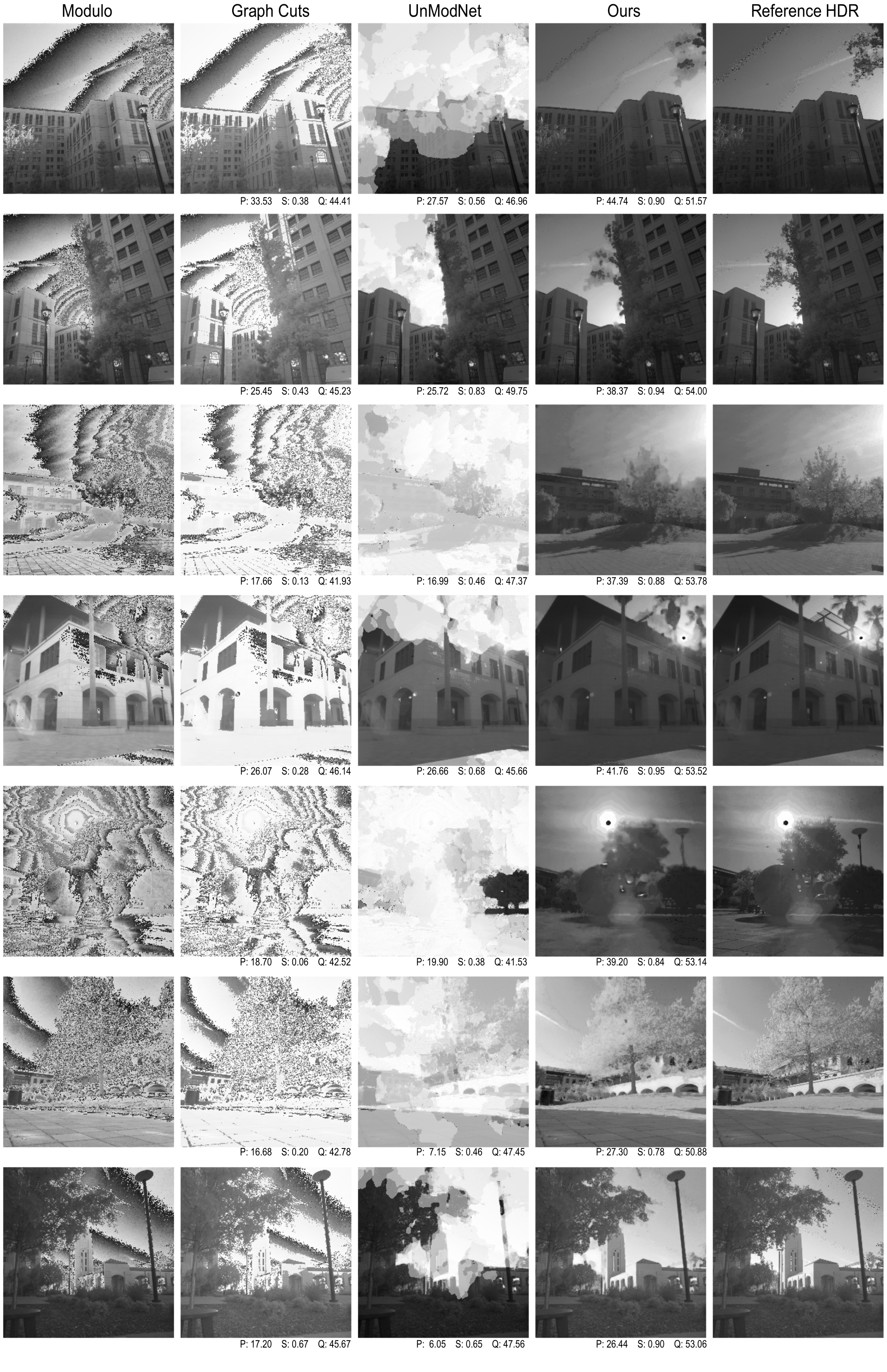}
\end{center}
   \caption{Comparisons on captured data.}
\label{fig:scamp_supplement1}
\end{figure*}

\begin{figure*}[t]
\begin{center}
   \includegraphics[width=0.8\linewidth]{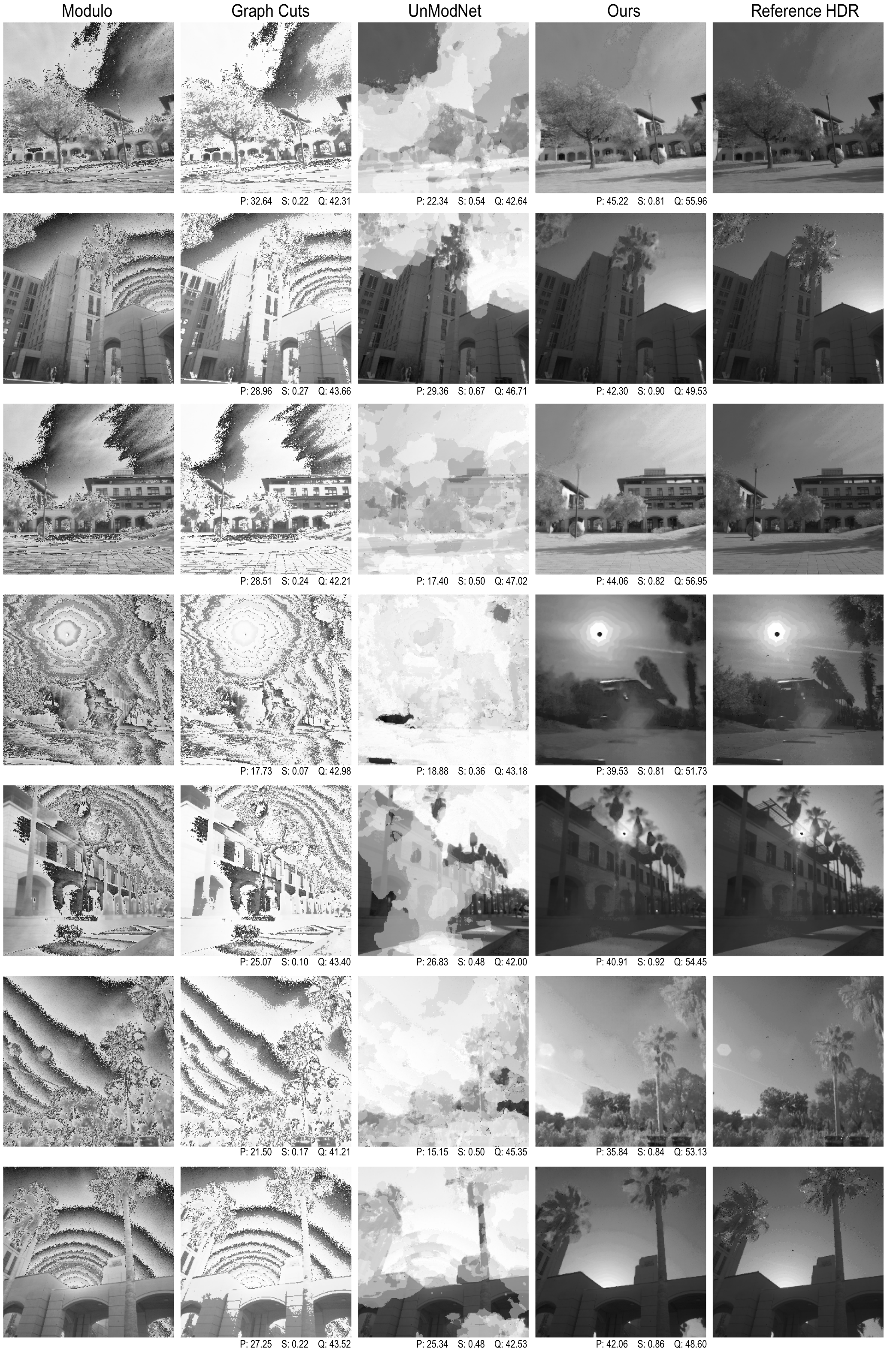}
\end{center}
   \caption{Comparisons on captured data.}
\label{fig:scamp_supplement3}
\end{figure*}

\begin{figure*}[t]
\begin{center}
   \includegraphics[width=0.8\linewidth]{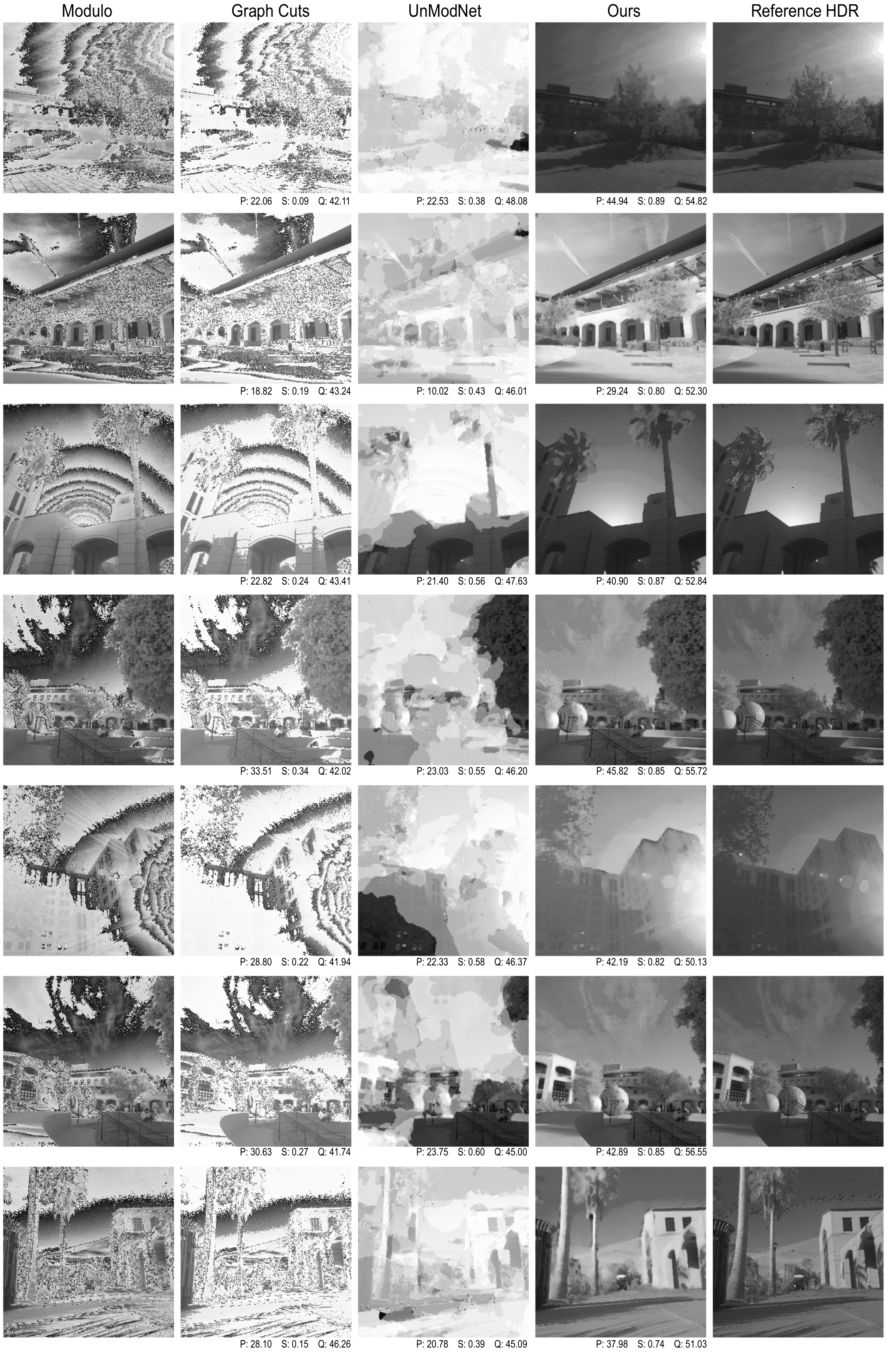}
\end{center}
   \caption{Comparisons on captured data.}
\label{fig:scamp_supplement4}
\end{figure*}

\begin{figure*}[t]
\begin{center}
   \includegraphics[width=0.8\linewidth]{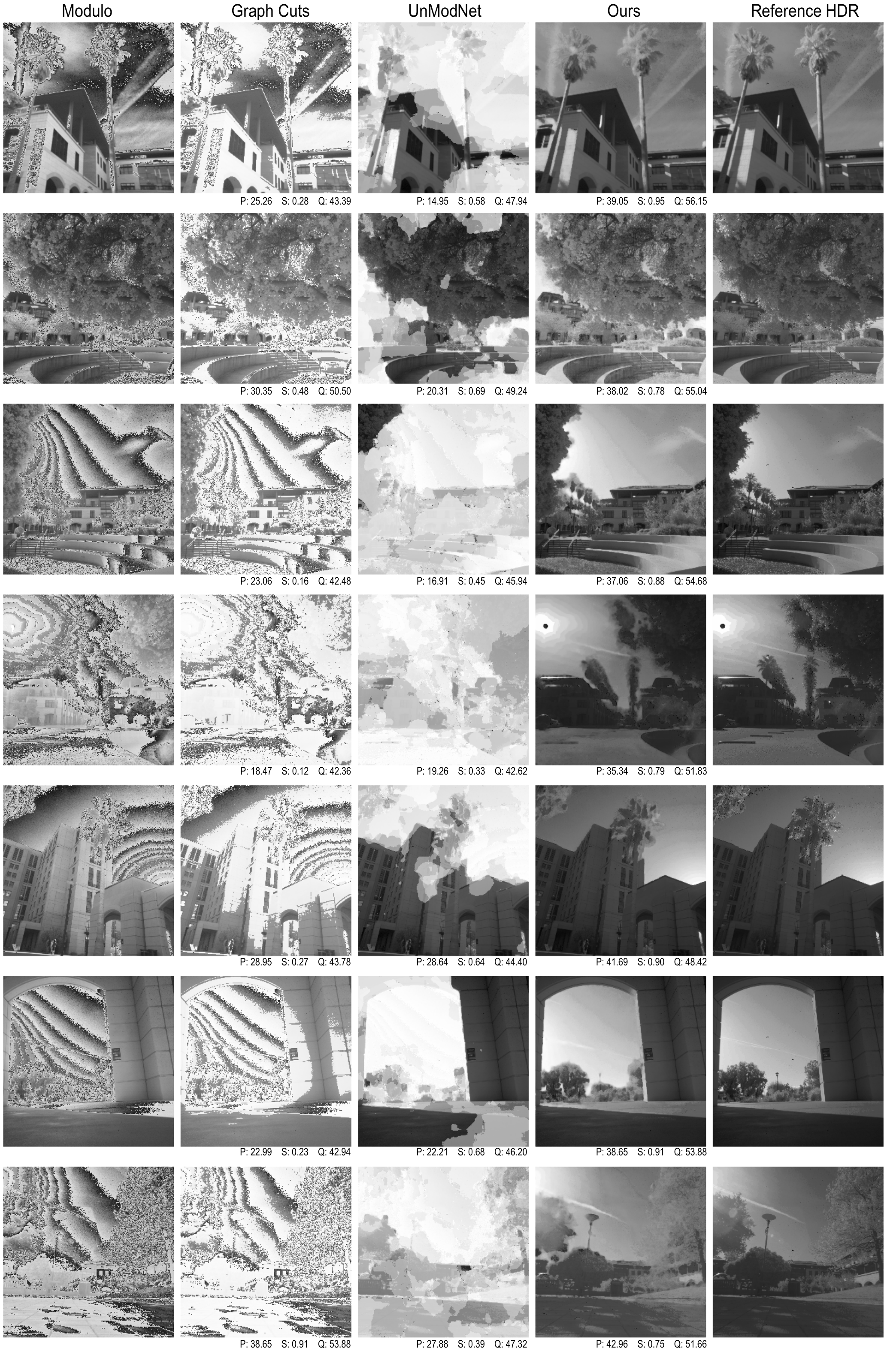}
\end{center}
   \caption{Comparisons on captured data.}
\label{fig:scamp_supplement5}
\end{figure*}

\begin{figure*}[t]
\begin{center}
   \includegraphics[width=0.8\linewidth]{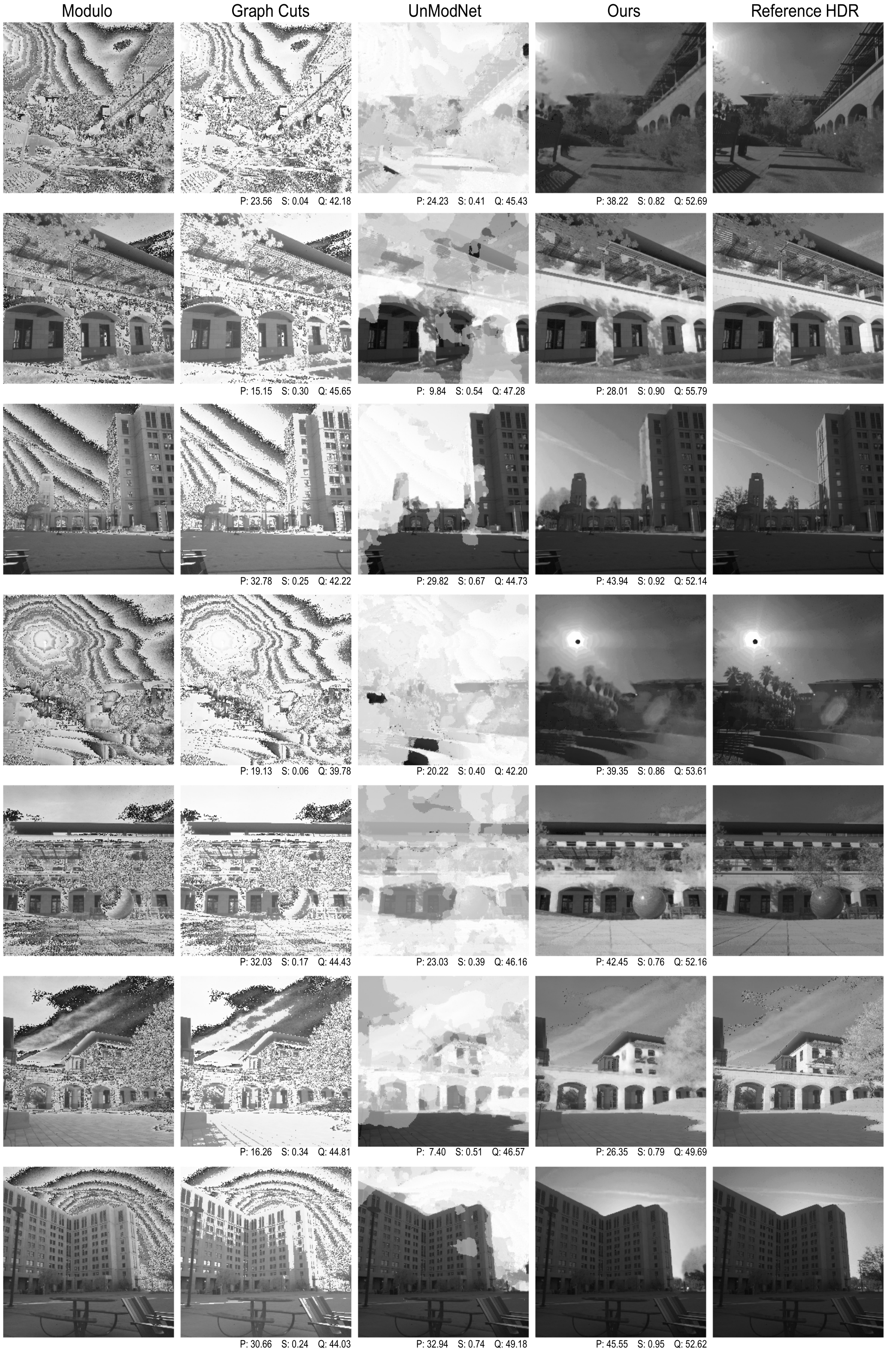}
\end{center}
   \caption{Comparisons on captured data.}
\label{fig:scamp_supplement6}
\end{figure*}

See Figures~\ref{fig:supplement1} and~\ref{fig:supplement2} for additional results. From left to right, each row shows the modulo image, the graph cuts method, UnModNet + modulo, Ours + modulo, the mantissa image, Ours + mantissa, and the ground truth image. All tonemapped images follow the tone-mapping described in Section~3. Additionally, we performed a study on the effects of noise. Our networks and comparisons were not trained on noisy images, so as we increase additive Gaussian noise, the PSNRs decrease, as shown in figure~\ref{fig:noise_study}. However, if the networks are trained with real data, they are able to capture the effects of noise, as demonstrated by the results from our reconstruction algorithm on the captured images with our prototype. Figures ~\ref{fig:scamp_supplement1}--\ref{fig:scamp_supplement6} show additional results for captured data with the SCAMP-5.

\end{multicols}
